\documentclass[longauth,oldversion]{aa} 
\usepackage{graphicx}
\usepackage{txfonts}
\usepackage{natbib}
\usepackage{color}
\bibpunct{(}{)}{;}{a}{}{,} 
\begin{document}
\title{Discovery of very-high-energy $\gamma$-ray emission from the
  vicinity of PSR~J1913+1011 with H.E.S.S.}

\author{F. Aharonian\inst{1,13}
 \and A.G.~Akhperjanian \inst{2}
 \and U.~Barres de Almeida \inst{8} \thanks{supported by CAPES Foundation, Ministry of Education of Brazil}
 \and A.R.~Bazer-Bachi \inst{3}
 \and B.~Behera \inst{14}
 \and M.~Beilicke \inst{4}
 \and W.~Benbow \inst{1}
 \and K.~Bernl\"ohr \inst{1,5}
 \and C.~Boisson \inst{6}
 \and O.~Bolz \inst{1}
% ...
}

\institute{
Max-Planck-Institut f\"ur Kernphysik, Heidelberg, Germany
\and
 Yerevan Physics Institute, Armenia
\and
Centre d'Etude Spatiale des Rayonnements, CNRS/UPS, Toulouse, France
\and
Universit\"at Hamburg, Institut f\"ur Experimentalphysik, Germany
\and
Institut f\"ur Physik, Humboldt-Universit\"at zu Berlin, Germany
\and
LUTH, Observatoire de Paris, CNRS, Universit\'e Paris VII, Meudon, 
France
\and
DAPNIA/DSM/CEA, CE Saclay, Gif-sur-Yvette, France
\and
University of Durham, Department of Physics, U.K.
\and
Unit for Space Physics, North-West University, Potchefstroom, South Africa
\and
Laboratoire Leprince-Ringuet, Ecole Polytechnique, CNRS/IN2P3, Palaiseau, France
\and 
Laboratoire d'Annecy-le-Vieux de Physique des Particules, CNRS/IN2P3,
Annecy-le-Vieux, France
\and
Astroparticule et Cosmologie (APC), CNRS, Universit\'e Paris VII,
Paris, France
\thanks{UMR 7164 (CNRS, Universit\'e Paris VII, CEA, Observatoire de Paris)}
\and
Dublin Institute for Advanced Studies, Ireland
\and
Landessternwarte, Universit\"at Heidelberg, K\"onigstuhl, Germany
% ...
}

\offprints{stefan.hoppe@mpi-hd.mpg.de, emmanuel.moulin@cea.fr}

\date{Received 21 September 2007 ; Accepted 20 Februar 2008}

\abstract{ The H.E.S.S. experiment, an array of four Imaging
  Atmospheric Cherenkov Telescopes with high sensitivity and large
  field-of-view, has been used to search for emitters of
  very-high-energy (VHE, $>$100 GeV) $\gamma$-rays along the Galactic
  plane, covering the region 30\degr \space $<\, l\, <$ 60\degr,
  280\degr \space $<\, l \,<$ 330\degr\, , and $-$3\degr \space $<\,
  b\, <$ 3\degr. In this continuation of the H.E.S.S. Galactic Plane
  Scan, a new extended VHE $\gamma$-ray source was discovered at
  $\alpha_{2000}$=19$^{h}$12$^{m}$49$^{s}$,
  $\delta_{2000}$=$+$10\degr09\arcmin06\arcsec (HESS~J1912$+$101).
  Its integral flux between 1-10\,TeV is $\sim$\,10\% of the Crab
  Nebula flux in the same energy range. The measured energy spectrum
  can be described by a power law $dN/dE \, \sim \, E^{-\Gamma}$ with
  a photon index $\Gamma$ = 2.7 $\pm$ 0.2$_{\mbox{stat}}$ $\pm$
  0.3$_{\mbox{sys}}$. HESS~J1912$+$101 is plausibly associated with
  the high spin-down luminosity pulsar PSR~J1913$+$1011. We also
  discuss associations with an as yet unconfirmed SNR candidate
  proposed from low frequency radio observation and/or with molecular
  clouds found in $^{13}$CO data.}
\keywords{Surveys - ISM:supernova remnants - ISM:individual
  objects:PSR~J1913$+$1001 - Gamma rays:observations - ISM:individual
  objects:HESS~J1912$+$101}
%
\maketitle
\section{Introduction}
\begin{figure*}[t!]
 \resizebox{\hsize}{!}{\includegraphics{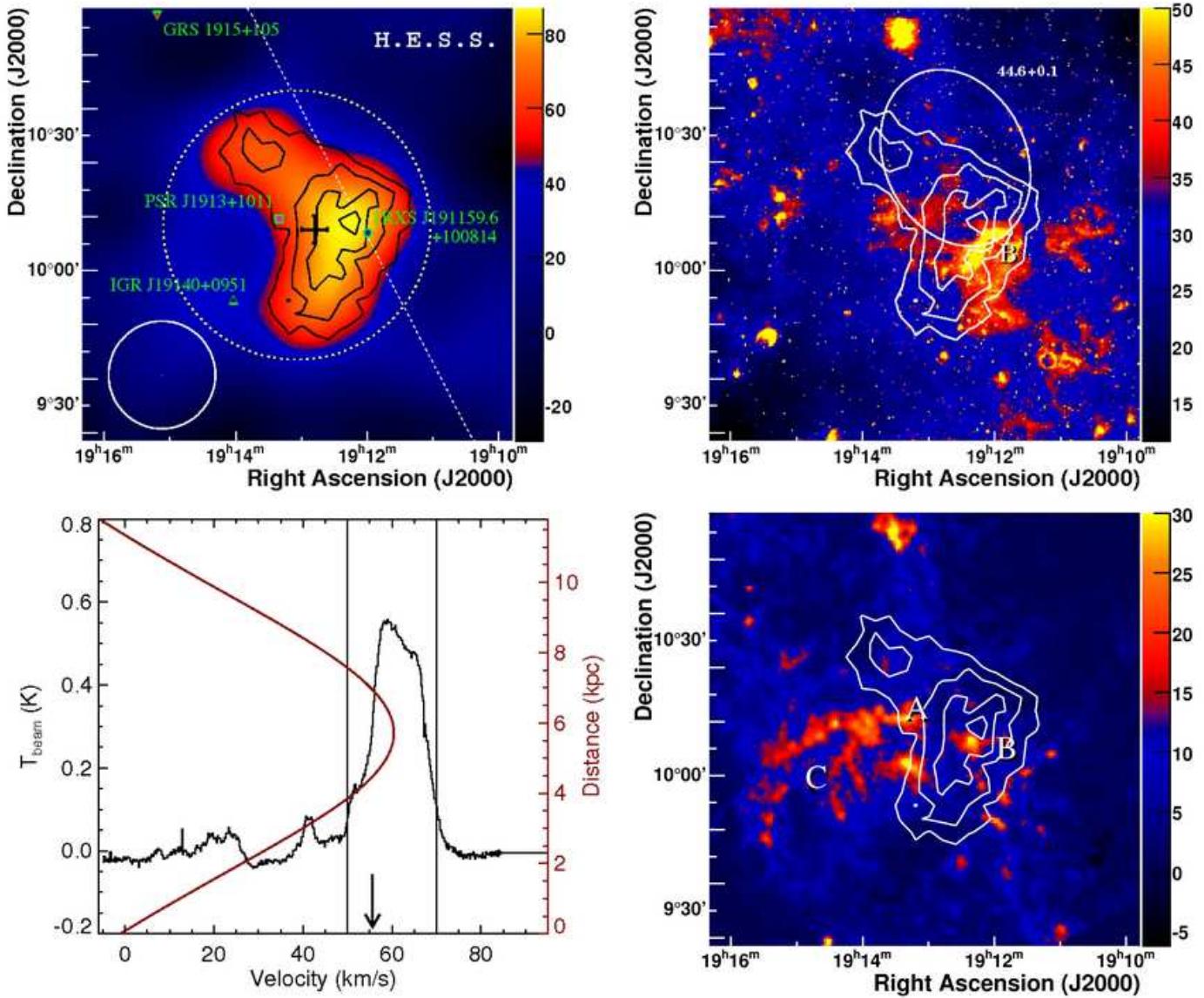}}
 \caption{{\it Top, left}: Image of the VHE $\gamma$-ray excess from
   HESS\,J1912$+$101, smoothed with a Gaussian profile of
   $\sigma$=0.13\degr \, along each axis. Significance contours
   calculated for an integration radius of
   $\theta_{\mbox{cor}}$=0.22\degr\space are shown in black at 3.5,
   4.5, 5.5 and 6.5 $\sigma$. The black cross indicates the best fit
   position of the source centroid together with its statistical
   errors. The dashed circle indicates the region from which the
   spectrum shown in Fig. \ref{fig2} has been extracted.  The
   positions of the pulsar PSR\,J1913$+$1011, the ROSAT source
   1RXS\,J191159.6$+$100814, the INTEGRAL source IGR\,J19140$+$0951
   and the microquasar GRS\,1915$+$105 are marked. The Galactic plane
   is indicated by a white dotted line. The white solid circle
   illustrates the 68\% containment radius of the Gaussian brightness
   profile used for smoothing convolved with the point spread function
   of the instrument.  {\it Top, right}: Image of the surface
   brightness (in units of MJy$\,$sr$^{-1}$) measured at $\sim$\,8.0
   $\mu$m within the Galactic Legacy Infrared Mid-Plane Survey
   Extraordinaire (GLIMPSE) using the Spitzer telescope
   \citep{GLIMPSE}. The white contours indicate the significance of
   the VHE $\gamma$-ray excess. The region marked as ``B'' is a
   complex of molecular clouds and HII-regions.  The white ellipse
   illustrates the position and extension of the SNR candidate
   44.6+0.1 proposed in the Clark Lake 30.9\,MHz Galactic plane survey
   \citep{kassim}. {\it Bottom, left}: Velocity profile of
   $^{13}$CO(J=1$\rightarrow$0) intensity at 110.2\,GHz from the
   Galactic Ring Survey \citep[GRS,][]{co13survey}, integrated within
   the dashed circle shown in the top, left figure. The velocity
   resolution is 0.25\,km\,s$^{-1}$. The distance/velocity
   correspondence from the Galactic rotation model from Fitch et
   al. \cite{galrotmodel} is also shown. The velocity corresponding to
   the nominal distance of PSR\,J1913$+$1011 (4.48\,kpc)
   \citep{pulsarcatalog} is marked by an arrow together with the velocity range of
   the image shown in the bottom, right figure. {\it Bottom, right}:
   GRS $^{13}$CO(J=1$\rightarrow$0) intensity in units of
   K$\,$km$\,$s$^{-1}$.  Intensities are integrated in the velocity
   range 50 - 70 km s$^{-1}$, corresponding to a distance of
   $\sim$\,5\,kpc.  The molecular cloud marked as 'A' is coincident with
   PSR\,J1913$+$1011. In region 'C' no significant VHE $\gamma$-ray
   excess was observed so far, even though it contains molecular
   clouds at a similar distance.}
\label{fig1}
\end{figure*}
A significant fraction of the recently discovered sources of
very-high-energy $\gamma$-rays (VHE; E $>$ 100\,GeV) within the
Galactic plane are associated with pulsar wind nebulae (PWNs) of high
spin-down luminosity pulsars, such as G\,18.0$-$0.7
\citep{hessj1825p2} and MSH~15$-$5{\it2} \citep{hessmsh1552}, or with
expanding shells of supernova remnants (SNRs) such as
RX\,J1713.7$-$3946 \citep{RXJ1713} and RX\,J0852.0$-$4622
\citep{VelaJr}. The observed VHE $\gamma$-ray emission from these
objects provides direct evidence for the existence of particles
accelerated to energies of at least $\sim$\,$10^{14}$ eV
\citep{RXJ1713}. In the case of accelerated electrons the VHE
$\gamma$-rays are produced via inverse-Compton scattering on ambient
photon fields, or, in case of dense environments, via non-thermal
bremsstrahlung. In the case of accelerated hadrons, $\pi^{0}$ mesons
and other hadrons are produced in interactions with nuclei in the
interstellar medium.  The $\gamma$-rays result from the decay of these
particles, in particular of the $\pi^{0}$s.\\
Most of the Galactic objects expected to emit VHE $\gamma$-rays are
related to some phase in the evolution of massive stars, either to
their wind outflows or to PWNs and SNR shells formed after their
collapse. Their spatial distribution is characterised by close
clustering along the Galactic plane \citep[rms in latitude
$\sim$\,0.3\degr,][]{hesssurveyI}. A systematic survey of the Galactic
plane is therefore an effective approach to discovering new
$\gamma$-ray sources and source classes. Due to its large
field-of-view, its high sensitivity and its location in the southern
hemisphere, the High Energy Stereoscopic System (H.E.S.S.)  is the VHE
instrument currently best-suited to performing such a survey.  The first
phase of the survey has resulted in numerous detections of VHE
$\gamma$-ray emitters in the inner Galactic region $|l|$ $<$
30\degr\space and $|b|$ $<$ 3\degr\space
\citep{hesssurveyI,hesssurveyscience}.  In 2005-2007 the survey has
been extended, covering the regions $l$ = 280\degr\space -
330\degr\space and $l$ = 30\degr\space - 60\degr\space.  Here we
present one of the new sources, HESS\,J1912$+$101, discovered during
these observations, which is a good candidate for a VHE PWN such as
HESS\,J1825$-$137 \citep{hessj1825p2}, HESS\,J1718$-$385 and
HESS\,J1809-193 \citep{hesstwopwn}.

\section{Observations and Results}
\begin{figure}
  \resizebox{\hsize}{!}{\includegraphics{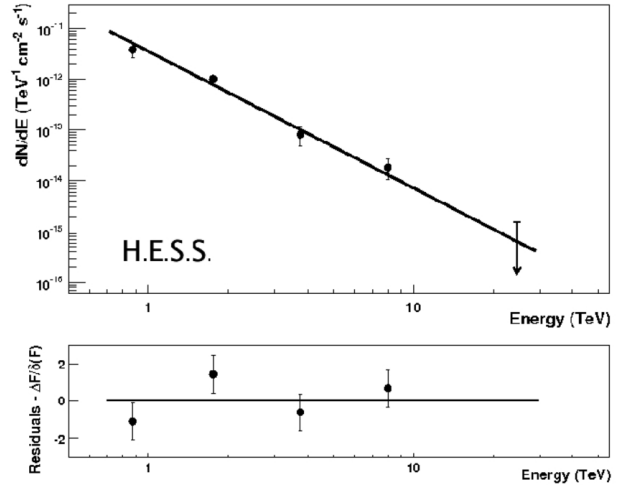}}
  \caption{Differential energy spectrum of HESS\,J1912$+$1011,
    extracted from the circular region indicated in Fig.\ref{fig1}
    (top, left). Events with energies between 0.55 and 55 TeV were
    used in the determination of the spectrum. They were binned with
    three bins per decade -- the resulting number of bins is 6 -- and
    then fitted with a power law (solid line). The two highest energy points of
    the spectrum were merged and are shown here as a 2$\sigma$ upper limit.}
  \label{fig2}
\end{figure}

The region of interest, which includes the microquasar GRS\,1915$+$105
\citep{GRS1915dist}, the Integral source IGR\,J19140$+$0951
\citep{integralid}, the high spin-down luminosity pulsar
PSR\,J1913$+$1011 \citep{ParkesII}, as well as the SNR candidate
44.6+0.1 \citep{kassim}, was first targeted in 2004 as part of the
observational programme on GRS\,1915$+$105. In 2005, dedicated
observations were taken on the supernova remnant W\,49B
\citep{W49Bhist}; some of the pointings also cover the field-of-view
of interest. In 2006, the region was observed again, as part of the
extended H.E.S.S. Galactic Plane Survey and with dedicated
observations of IGR\,J19140$+$0951. PSR\,J1913$+$1011 was selected for
dedicated observations as part of a systematic study of high spin-down
luminosity pulsars, but was not observed due to the planned coverage
of this region in other programmes. The combined data set was
investigated using the standard survey analysis (a radius of the
on-source region of $\theta_{\mbox{cut}}$=0.22\degr, a ring background
region of radius 0.8\degr\, and \emph{hard} cuts, which include a
minimum requirement of 200 photo electrons per shower image, for
$\gamma$-ray selection) as described in Aharonian et al.
\cite{hesssurveyI_year}.\\
The analysis presented considers only observations pointed within
3\degr\, of the original source candidate apparent within the Galactic
plane survey analysis. The mean pointing offset is 1.1\degr. After
quality selection to remove data affected by unstable weather
conditions or hardware issues, the data set has an acceptance
corrected live time (equivalent time spent at an offset of 0.5\degr)
of 20.8\,h at the centre of the emission. The zenith angles of the
observations range from 33\degr\, to 54\degr\, leading to a typical
energy threshold of $\sim$\,800 GeV. Figure \ref{fig1}(top, left)
shows the excess count map of the 1.6\degr\,$\times$\,1.6\degr\space
region around the source smoothed with a Gaussian profile of width
0.13\degr\space to reduce statistical fluctuations. A clear excess of
VHE $\gamma$-rays is observed with a peak statistical significance of
7.5 $\sigma$ for an integration radius of
$\theta_{\mbox{cut}}$=0.22\degr. As the source was discovered as part
of the H.E.S.S. Galactic Plane Survey, the statistical trials
associated with this survey must be taken into account
\citep[see][]{hesssurveyI}.  A very conservative estimate of the
number of trials which accounts for the large number of sky positions
probed ($\sim$\,5$\times$10$^{5}$) leads to a post-trials significance
of 5.5 $\sigma$.  Fitting the uncorrelated excess count map with a
symmetric Gaussian profile convolved with the point spread function of
the instrument leads to a best fit position of
$\alpha_{2000}$=19$^{h}$12$^{m}$49$^{s}$,
$\delta_{2000}$=$+$10\degr09\arcmin6\arcsec, with a 3\arcmin\space
statistical error in each coordinate, as indicated by the black cross
in Fig. \ref{fig1} (top, left).  HESS\,J1912$+$101 is clearly extended
with an intrinsic Gaussian width of $\approx$ 0.26\degr \, $\pm$
0.03\degr$_{\mbox{stat}}$, with a fit probability for a Gaussian
source profile of 9\%. Ellipsoidal fits do not improve the fit
probability. No further statements about the morphology are possible
at the level of statistics available at the moment.
\\
For the spectral analysis only observations pointed within
2.0\degr\space of the centre of the source were used, to remove events
with large uncertainties in reconstructed energy. The live time of the
remaining dataset is 18.7 h. The spectrum was determined within a
circular region of 0.5\degr\, radius (indicated by a dashed circle in
Fig.\ref{fig1}(top, left)), which represents an $\sim$\,90\% enclosure
of the excess, chosen as a compromise between optimal signal to noise
ratio and independence of source morphology. To minimise spectral
uncertainties, the background was estimated from regions with equal
offset from the centre of the field-of-view as described in Berge et
al. \cite{hessbackground}. The dataset contains a large fraction of
observations targeted at other possible sources within the
field-of-view (e.g. GRS\,1915$+$105) and was taken over a time period
of three years. Furthermore, due to the source extension, the
observations of HESS\,J1912$+$101 have rather uneven exposure across
the region and coverage over time. Consideration of only nearby
H.E.S.S. pointing positions, use of the reflected-region background
model and the estimation of the time-dependent optical response of the
system, corrects for these variations. However, these factors lead to
somewhat increased systematic errors with respect to typical
H.E.S.S. results.  Within the large integration circle 276 excess
events were found, corresponding to a statistical significance of 7.5
$\sigma$ (pre-trials) for this dataset used for spectral analysis. The
spectrum shown in Fig. \ref{fig2} can be described by a power law:
$\mathrm{d}N/\mathrm{d}E =
\Phi_{\mathrm{0}}(E/1\mathrm{TeV})^{-\Gamma}$ with photon index
$\Gamma=2.7\pm0.2_{\mathrm{stat}}\pm0.3_{\mathrm{sys}}$ and flux
normalisation
$\Phi_{\mathrm{0}}=(3.5\pm0.6_{\mathrm{stat}}\pm1.0_{\mathrm{sys}})
\times 10^{-12}$\space cm$^{-2}$ s$^{-1}$ TeV$^{-1}$ . The fit has a
$\chi^2$/ndf = 6.3/4. The integral flux between 1 and 10 TeV is about
9\% of the flux of the Crab Nebula in the same energy range
\citep{hesscrab}.
\section{Possible Associations}
The celestial region around HESS\,J1912$+$101 hosts several potential
counterparts.  While a superposition of two or more is also possible,
each of the following objects could individually account for the
observed
$\gamma$-ray emission:\\
The most plausible counterpart candidate, both in positional and
energetic connection to HESS\,J1912+101 is the pulsar {\bf
  PSR\,J1913$+$1011} \citep{ParkesII}, which is slightly offset from
the H.E.S.S. source's best fit position by $\sim$\,0.15\degr,
spatially consistent at the $\sim$\,3$\sigma$ level.
It is a rather energetic pulsar with a spin-down luminosity of
2.9$\times$10$^{36}$ erg s$^{-1}$, a spin-down age of
$\tau_{\mathrm{c}}\space \approx \space$ $1.7 \space\times\space 10^5$
years, a spin period of 36 ms and a distance estimated from dispersion
measurements of 4.48\,kpc \citep{pulsarcatalog}.  The pulsar is
sufficiently energetic to power the H.E.S.S. source with an implied
conversion efficiency from rotational energy to 1-10 TeV $\gamma$-rays
of $\epsilon$=0.5\%, comparable to the efficiency inferred for other
VHE PWN candidates such as HESS\,J1718$-$385 and HESS\,J1809$-$193
\citep{hesstwopwn}. Its distance would result in a projected size of
HESS\,J1912$+$101 of
$\sim$\,70\,pc.\\
These characteristics suggest an association of the VHE emission with
the pulsar's wind nebula, similar to other PWN associations such as
Vela\,X \citep{hessvelax}, HESS\,J1825$-$137 \citep{hessj1825p2} and
MSH\,15$-$5{\it2} \citep{hessmsh1552}. Compared with these examples,
HESS\,J1912+101 would be the oldest candidate for a VHE emitting PWN.
In leptonic models of VHE $\gamma$-ray production, a PWN is expected
to also emit non-thermal X-rays. Two Chandra observations within the
region of interest are currently available. The first (Obs. Id 4590)
was targeted on EXO\,1912$+$097 \citep{EXO1912}, which is coincident
with and possibly the same object as IGR\,J19140$+$0951
\citep{integralid}. The overlap of this observation with
HESS\,J1912$+$101 is marginal. The second (Obs. Id 3854) was targeted
on the pulsar PSR\,J1913$+$1011 and covers a larger fraction of the
H.E.S.S. source.  No X-ray PWN was reported at the position of
PSR\,J1913$+$1011 \citep[see][]{chandrapwn}. Existing Chandra
observations do not allow for a counterpart search on the same spatial
scale as the H.E.S.S. source, and a detailed analysis of the Chandra
data is beyond the scope of this paper.  Despite considerably less
sensitivity, the ASCA satellite with its larger field-of-view seems
more suitable for counterpart studies on larger scales. Two ASCA
observations coincide with HESS\,J1912$+$101. The first (Obs. Id
57005060) covers the whole H.E.S.S. source but suffers strongly from
stray light contamination, the second (Obs. Id 57005070) barely covers
the peak of the VHE
emission. Therefore no ASCA data has been analysed for this paper.\\
In the emerging picture of PWN associations the VHE emission peak is
often found offset from the pulsar position (e.g. HESS\,J1825$-$137,
MSH\,15$-$5{\it2}, HESS\,J1718$-$385 and HESS\,J1809$-$193). These
offsets can be explained in terms of the proper motion of the pulsar
and/or the expansion of the SNR/PWN into an inhomogeneous medium
\citep[see e.g.][]{pwnasym}. In the case of HESS\,J1912$+$101 the
observed offset could be explained by proper motion assuming a
velocity of $\sim$\,60 km s$^{-1}$ (to travel the $\sim$\,10 pc from
the centroid of the VHE emission to the current pulsar location in
1.7$\space\times\space$10$^5$ years). The hypothesis that the PWN of
PSR\,J1913$+$1011 (and the associated SNR) is expanding into an
inhomogeneous medium seems quite plausible as the
$^{13}$CO(J=1$\rightarrow$0) emission line measurements taken at 110.2
GHz indicate the existence of `clumpy' molecular material (in
particular the area ``A'' as indicated in Fig. \ref{fig1}(bottom,
right)) at the pulsar position and roughly at its distance. Such a
scenario for an asymmetric nebula was also proposed for
HESS~J1825$-$137 \cite{hessj1825p2}. HESS~J1912$+$101 and
HESS~J1825$-$137 share several similarities: 1) similar offset of the
VHE emission region from the pulsar position; 2) similar angular size
(rms size of 0.24\degr \space for HESS~J1825$-$137 and 0.26\degr
\space for HESS~J1912$+$101); 3.) similar distances (inferred from
dispersion measurements of the corresponding pulsars to be
$\sim$\,4\,kpc). The intrinsic sizes are hence both $\sim$\,40\,pc.
In a leptonic scenario, the cooling timescale, in combination with the
propagation speed of the lowest energy ( and hence longest-lived)
electrons which upscattering soft photons to energies detectable by
H.E.S.S., limit the extent of the emission region.  As argued by
Aharonian et al. (2006d) synchrotron cooling is likely to be the
dominant energy-loss process for magnetic field strengths
$\mathrm{B}$\,$>$\,3\,$\mu$G and multi-TeV electrons. For scattering
in the Thomson regime by cosmic microwave background radiation (CMBR)
photons, the parent electron population responsible for the VHE
$\gamma$-rays detected by H.E.S.S. have energies 10--100 TeV. The
magnetic field strength within the extended X-ray PWN of
PSR~B1823$-$13 suggested by Gaensler et al. \cite{gaensler1823} is
$\sim$\,10\,$\mu$G. Considering the different sizes of the X-ray and
VHE $\gamma$-ray emission regions (5' versus 60') for PSR~B1823$-$13,
the magnetic field strength in the outer VHE nebula should be lower
than in the more compact X-ray nebula \citep[see][]{okkiepwn}. Here we
assume similar magnetic field strengths for HESS~J1825$-$137 and
HESS~J1912$+$101. The corresponding lifetime due to synchrotron losses
of electrons upscattering CMBR photons to energies $E_{\gamma}$ is
given by
  \begin{equation}
    \tau_{\mathrm{synch}} \approx 3 \times 10^{4}\:
    \Big(\frac{\mathrm{B}}{5\mu\mathrm{G}}\Big)^{-2}\Big(\frac{\mathrm{E}_{\gamma}}{\mathrm{TeV}}\Big)^{-1/2}
    \mathrm{years .}
  \end{equation}
  Even for a magnetic field strength of 5\,$\mu$G -- a value similar
  to the interstellar magnetic field -- and 1\,TeV $\gamma$-rays is
  the synchrotron cooling time scale for the upscattering electrons a
  factor $\sim$\,6 lower than the age of PSR~J1913$+$1011
  ($\tau_{c}=1.7 \times 10^{5}$ years) and still close to the age of
  PSR~B1823$-$13 ($\tau_{c}=2.1 \times 10^{4}$ years). This provides a
  natural explanation for the similar sizes of HESS~J1825$-$137 and
  HESS~J1912$+$101 even though the spin-down ages of the corresponding
  pulsars are different by a factor of $\sim$\,8. Assuming the
  transport process to be dominantly advective rather than diffusive,
  the necessary average flow speed needed to drive electrons from the
  position of the pulsar to the edge of the $\gamma$-ray nebula
  ($\mathrm{D}^{\mathrm{trav}}\approx
  40\frac{\mathrm{d}}{4.5\mathrm{kpc}} \mathrm{pc} $) within the
  synchrotron loss time is
\begin{equation}
v = \frac{\mathrm{D}^{\mathrm{trav}}}{\tau_{\mathrm{synch}}} \approx 1400
\: \Big(\frac{\mathrm{d}}{4.5\mathrm{kpc}}\Big)\Big(\frac{\mathrm{B}}{5\mu\mathrm{G}}\Big)^{2}\Big(\frac{\mathrm{E}_{\gamma}}{\mathrm{TeV}}\Big)^{1/2}
\:\mathrm{km}\mathrm{s}^{-1} \:\: .
\label{speed}
\end{equation}
This speed is fairly large, but may not be implausible if the
associated SNR expanded at least in part into the low-density phase of
the interstellar medium. Assuming an energy-independent advection
speed, equation \ref{speed} predicts an energy-dependent morphology
for the nebula, as was observed for HESS~J1825$-$137, where the photon
index changes with distance by $\Delta \: \Gamma \: \approx 0.6$
\cite{hessj1825p2}. With the available statistics it is not possible
to extract position-resolved spectra for HESS~J1912$+$101 to test this
hypothesis also in this case. We also note that a possibly significant
fraction of the observed VHE emission could also result from IC
scattering of higher energy target photons, e.g. emission from the
close-by molecular clouds and HII-regions as visible in Fig.\ref{fig1}
(top, right). This component of the VHE emission would probe lower
energy electrons with significantly higher cooling times. Its size
would therefore be limited
not by the cooling time but by the age of the system.\\
Whilst the spin-down powers of PSR~B1823$-$13 and PSR~J1913$+$1011 are
very similar ($\sim$\,3$\times$10$^{36}$ erg s$^{-1}$),
HESS~J1825$-$137 is a more luminous source than HESS~J1912$+$101,
resulting in fairly different apparent VHE conversion efficiencies of
3\% for the former and 0.5\% for the latter. The spectral indices of
the VHE sources are, however, compatible within statistical
uncertainties
($\Gamma=2.38\pm0.02_{\mathrm{stat}}\pm0.15_{\mathrm{sys}}$ and
$\Gamma=2.7\pm0.2_{\mathrm{stat}}\pm0.3_{\mathrm{sys}}$).  Assuming a
continous injection of electrons with a power law $dN/dE \approx
E^{-\alpha}$ where $\alpha$ = 2.0 .. 2.2, $\alpha$ is increased to 3.0
... 3.2 by radiative cooling, and a VHE $\gamma$-ray spectrum with a
photon index of 2.0 to 2.1 is expected above a synchrotron break
energy of $\sim$ 30\,GeV for HESS~J1912+101. 
In Aharonian et al. \cite{hessj1825p1} different scenarios were
discussed to explain the observed softer photon index, including the
effect of Klein-Nishina corrections to the IC scattering cross-section
and the superposition of IC spectra arising from electrons injected in
the past. Both these arguments are equally
applicable to HESS~J1912+101.\\
Another candidate object for the observed emission is the shell of a
hypothetical {\bf supernova remnant} which would be rather old if
associated with the birth of PSR\,J1913$+$1011 (spin-down age $1.7
\times 10^{5}$ years). In this scenario, the observed $\gamma$-ray
emission could be explained as the result of the interaction of
accelerated particles from the SNR with dense molecular clouds. A
candidate for a supernova remnant assocation is 44.6+0.1, an yet
unconfirmed SNR candidate proposed in the Clark Lake 30.9\,MHz
Galactic plane survey \citep{kassim} and further investigated by
Gorham \cite{gorham}. The estimated position and size of this object
are illustrated by a white ellipse in Fig.1 (top,right). 44.6+0.1
overlaps with the VHE emission region, but the overall match in
morphology is rather poor. However, the resolution from the Clark
Lake survey is insufficient for a firm conclusion about the morphology
and only part of the remnant might be visible at 30.9\,MHz. No
SNR has been detected in X-rays. The only published keV X-ray sources
in this region are the faint ROSAT sources 1RXS\,J191159.6$+$100814,
1RXS\,J191254.3$+$103807 and 1RXS\,J191114.8$+$102102
\citep{rosatfaint}. Such a lack of X-ray emission could be explained
using for example the model of Yamazaki et al \cite{oldsnrtev}, in
which an age of the SNR of 1.7$\space\times\space$10$^5$ years would
imply a rather low X-ray to TeV energy flux ratio of $\sim$\,0.01 (for
the energy ranges 2-10 keV and 1-10 TeV).  In this scenario the
spatial distribution of the VHE emission should then trace the
distribution of the molecular clouds which act as target
material. Such clouds have been observed in the vicinity of
PSR\,J1913$+$1011 (indicated as area ``B'' in Fig.\ref{fig1} (top,
right)) in infrared at $\sim$\,8\,$\mu \mathrm{m}$ (Fig.\ref{fig1}
(top, right)), in the $^{13}$CO measurements presented above
(Fig.\ref{fig1} (bottom, right)) and in the GRS CS(J=2$\rightarrow$1)
map of the region. The best distance estimate for G44.3$+$0.1, an HII
region located within this region and coincident with
1RXS\,J191159.6$+$100814, is 3.9\,kpc \citep[from HI absorption
measurements,][]{distanceshII}, which -- given the uncertainties of
the two methods -- would place it at a distance consistent with that
of the pulsar (4.48\,kpc, from dispersion measurements). The VHE
emission peaks close to this dense region. There is, however, no
evidence for an overall spatial correlation of the $\gamma$-ray
emission with molecular clouds (see e.g. area ``C'' in Fig.\ref{fig1}
(bottom,right)).  In summary, we find little evidence to support the
scenario of an old SNR causing the observed VHE $\gamma$-ray signal, but
can not exclude this hypothesis on the basis of the available
experimental data.\\
{\bf Other celestial objects} in the vicinity of HESS\,J1912$+$101
include the microquasar GRS\,1915$+$105 and the Integral source
IGR\,J19140$+$0951. GRS\,1915$+$105 is $\sim$\,1\degr\,away from the
centroid of the VHE emission.
The suggested distances to GRS\,1915$+$105 (12.5\,kpc
\citep{GRS1915dist} and 6.5\,kpc \citep{kaiser}) would correspond
to a projected distance to the centroid of HESS\,J1912$+$101
of 110 to 220\,pc and the suggested jet
termination sites (IRAS\,19132$+$1035, IRAS\,19124$+$1106) are both
situated outside the VHE emission region.
A possible association of
GRS\,1915$+$105 and HESS\,J1912$+$101 is therefore considered as
unlikely. The relative position of IGR\,J19140$+$0951, which is
probably a Super Giant X-ray binary \citep{integralidatel}, renders a
dominant contribution to the VHE emission from this object unlikely as
well.
\section{Summary}
The continuation of the H.E.S.S. survey of the Galactic plane has
resulted in the discovery of a new extended VHE $\gamma$-ray source,
HESS\,J1912$+$101. The complexity of the vicinity of HESSJ\,1912+101
at other wavebands does not allow us to make a firm conclusion on the
nature of the source yet. The most plausible scenario appears to be
that the observed emission is associated with the high spin-down
luminosity pulsar PSR\,J1913$+$1011, which would make HESS~J1912$+$101
a $\gamma$-ray PWN similar to HESS~J1825$-$137 in both the characteristics of
the observed $\gamma$-ray emission and the energetics of the associated
pulsar. Ultimately, an energy-dependent morphology change needs to be
established in further H.E.S.S. observations to confirm the similarity
with HESS~J1825$-$137 as argued here. Also the PWN needs to be
detected in deep X-ray observations. An alternative possibility is the interaction of
accelerated particles from an old SNR with molecular material in a
dense region in the interstellar medium. Neither of the two scenarios
is fully supported by existing multiwavelength data, but the existence
of an energetic pulsar within the VHE emission region favours an
association with a PWN, although the confirmation of a PWN at other
wavebands is currently pending.

\begin{acknowledgements}
  The support of the Namibian authorities and of the University of
  Namibia in facilitating the construction and operation of
  H.E.S.S. is gratefully acknowledged, as is the support by the German
  Ministry for Education and Research (BMBF), the Max Planck Society,
  the French Ministry for Research, the CNRS-IN2P3 and the
  Astroparticle Interdisciplinary Programme of the CNRS, the
  U.K. Science and Technology Facilities Council (STFC), the IPNP of
  the Charles University, the Polish Ministry of Science and Higher
  Education, the South African Department of Science and Technology
  and National Research Foundation, and by the University of
  Namibia. We appreciate the excellent work of the technical support
  staff in Berlin, Durham, Hamburg, Heidelberg, Palaiseau, Paris,
  Saclay, and in Namibia in the construction and operation of the
  equipment.
  \\
  This publication makes use of molecular line data from the Boston
  University-FCRAO Galactic Ring Survey (GRS). The GRS is a joint
  project of Boston University and Five College Radio Astronomy
  Observatory, funded by the National Science Foundation under grants
  AST-9800334, AST-0098562, AST-0100793, AST-0228993, \& AST-05076
\end{acknowledgements}

\end{document}